\documentstyle[preprint,aps]{revtex}
\begin{document}
\draft
\title{ Integrable Systems on Flag Manifold  and\\
Coherent State Path Integral}

\author{Myung-Ho Kim\thanks{E-mail
 address:mhkim@yurim.skku.ac.kr}}
\address{ Department of Mathematics,
 Sung Kyun Kwan University,
 Suwon 440-746, KOREA }

\author{Phillial Oh\thanks{E-mail
 address:ploh@yurim.skku.ac.kr}}
\address{Department of Physics,
Sung Kyun Kwan University,
Suwon 440-746,  KOREA}

\maketitle

\begin{abstract}
We construct integrable models on flag manifold
by using the symplectic structure explicitly
given in the Bruhat
coordinatization of flag manifold.
They are non-commutative integrable
and some of the conserved quantities are given by the
 Casimir invariants.
We quantize the systems using the coherent state
path integral technique
and find the exact expression for the propagator
for some special cases.
\end{abstract}

\newpage

It is well known that there is a natural symplectic structure
on the coadjoint orbits of Lie group \cite{kiri} and it can be
used to define the generalized Poisson bracket
to describe Hamiltonian systems on the coadjoint
orbits \cite{arno}.
In fact, there also exists a natural Lie group action on
them and
they are equipped with Poisson-Lie bracket \cite{arno}.
The geometrical construction of completely integrable
systems
on such symplectic manifolds
is very interesting in view of the recent developments
in the theory of finite-dimensional integrable
models \cite{intg}.
In this paper, we construct integrable models on flag manifold
which is the maximal coadjoint orbit of $SU(N)$ group and
quantize the
system using the coherent state path
integral technique \cite{klau}.
Physically, these systems correspond to $SU(N)$ spin models
and they provide another method of
classical formulation of non-Abelian
Chern-Simons particles \cite{oh93}
and give the framework of geometric quantization for them.

We start with a brief summary of flag manifold which
is essential for the presentation.
More details can be found in Refs.\cite{helg} and \cite{press}.
$SU(N)$ flag manifold $M_N$ is defined as a set ${\cal S}$
of all nested sequence of linear subspaces of  the standard
complex $N$-space ${\bf C}^{N}$,
\begin{equation}
\{ (C_1,C_2,\cdots, C_{N-1})\vert \mbox{dim}
\ C_l=l, \ \ C_l\subset C_{l+1}\subset {\bf C}^N\}.
\end{equation}
Then, there is a natural $SU(N)$ group action on  ${\cal S}$
which is transitive. Let $x_0=({\bf C}^1,
{\bf C}^2,\cdots, {\bf C}^{N-1})\in
{\cal S}$. Then, we can see easily that  ${\cal S}=
M_N=SU(N)/T^{N-1}$,
where the maximal torus group
$T^{N-1}$ is the stabilizer group at $x_0$.
Let $SU_c(N)$ be the complexification
of $SU(N)$ and $B_N$ a Borel
subgroup of $SU_c(N)$. Then using the Iwasawa decomposition
$SU_c(N)=SU(N)B_N$, we have isomorphism $M_N=SU_c(N)/B_N$.
{}From this complex representation of $M_N$, one can prove that
$M_N$ is a complex manifold. Also, there is a natural
isomorphism of $M_N=SU(N)/T^{N-1}$ with the coadjoint
orbit $\{g\lambda g^{-1}\vert g\in SU(N)\}$ defined by
\begin{equation}
gT^{N-1}\rightarrow   g\lambda g^{-1}
\end{equation}
where $\lambda$ is an element of the dual
Lie algebra of $T^{N-1}$.
Hence, $M_N$ has the symplectic structure inherited
from the coadjoint orbit \cite{kiri}. Together with
the complex structure, $M_N$ becomes a K\"ahler manifold.

In this paper, we will concentrate on $M_3$ to make
the presentation simple. Analysis of higher $M_N$ will
appear elsewhere. In order to
 construct integrable models  explicitly, we have
to coordinatize the flag manifold $M_3$. The ideal choice
for the explicit  construction of the symplectic structure
seems to be the Bruhat coordinatization \cite{pick}.
 According to Bruhat cell decomposition\cite{helg},
the flag manifold $M_3$ can be covered with six
coordinate patches. The convenient thing about
Bruhat cell decomposition is that the largest
cell provides a coordinatization $(z_1,z_2,z_3)$
of nearly all of the flag manifold missing only
lower-dimensional
subspaces. Also on this largest cell,
the K\"ahler metric $ds^2=\sum_{i,j}g_{ij}dz_id\bar z_j$
and the corresponding symplectic structure
\begin{equation}
\Omega=i\sum_{i,j}g_{ij}dz_i\wedge d\bar z_j
\end{equation}
can be calculated \cite{pick}.

The largest cell on $M_3=SU_c(3)/B_3$ is
represented as follows:
\begin{equation}
[g_c(z)]_{B_3}=\left[\left(\begin{array}{ccc}
1&0&0\\
z_1&1&0\\
z_2&z_3&1\\
\end{array}
\right)\right]_{B_3}\mapsto (z_1,z_2,z_3)
\end{equation}
with $g_c\in SU_c(3)$.
Symplectic structure is given
by the K\"ahler potential $W$
which was calculated explicitly
by the holomorphic line bundle approach\cite{pick}
with the result
\begin{equation}
W=\log (1+\vert z_1\vert^2+\vert z_2\vert^2)^m
(1+\vert z_3\vert^2+\vert z_2-z_1z_3\vert^2)^n
\label{symp}
\end{equation}
where $m,n$ are integers.
Using the symplectic structure expressed in terms of $W$,
\begin{equation}
\Omega=i\partial\bar\partial W,
\end{equation}
we obtain the K\"ahler metric given by
\begin{equation}
g_{ij}=\frac{\partial^2W}{\partial z_i\partial\bar z_j}.
\end{equation}
Then the Poisson bracket is defined via
\begin{equation}
\{F,G\}=-i\sum_{i,k}g^{ki}\left(\frac{\partial
F}{\partial  z_k}\frac{\partial G}{\partial\bar z_i}-
\frac{\partial G}{\partial z_k}
\frac{\partial F}{\partial\bar z_i}\right)
\end{equation}
where the inverse metric $g^{ki}$ satisfies
$g_{ik}g^{kj}=\delta_i^j$.

Using the above symplectic structure, we
calculate the Hamiltonian function $F_a$ associated with
 the generator $T_a$
satisfying $[T_a, T_b]=-f_{ab}^c T_c$
 with the structure constants given in
the  Gell-Mann basis\cite{corn}. We first perform
 this calculation
for $T_3$ which is given by
\begin{equation}
T_3=\frac{1}{2}\left(\begin{array}{ccc}
i&0&0\\
0&-i&0\\
0&0&0\\
\end{array}
\right).
\end{equation}
Recall the definition of
the Hamiltonian vector field $X_3$ associated with $T_3$:
\begin{equation}
X_3=\frac{d}{dt}{\Big\vert}_{t=0}(\exp tT_3)\circ [g_c(z)]_B.
\end{equation}
Now, by multiplying a suitable $b\in B$, we have
\begin{equation}
(\exp tT_3)\circ [g_c(z)]_B= [g_c(z^\prime)]_B=
\left[\left(\begin{array}{ccc}
1&0&0\\
z_1^\prime&1&0\\
z_2^\prime&z_3^\prime&1\\
\end{array}
\right)\right]_B.
\end{equation}
A simple calculation shows
\begin{equation}
z_1^\prime=(1-it)z_1, \ \ z_2^\prime=(1-i\frac{t}{2})z_2,
 \ \ z_3^\prime=(1+i\frac{t}{2})z_3,
\end{equation}
from which we obtain
\begin{equation}
X_3=-\frac{i}{2}(2z_1\frac{\partial}{\partial z_1}+
z_2\frac{\partial}
{\partial z_2}-z_3\frac{\partial}{\partial z_3})+
\mbox{c.c}.
\end{equation}
The Hamiltonian function associated with the vector
field $X_3$ is
defined by \cite{arno}
\begin{equation}
X_3\rfloor\Omega = dF_3.
\end{equation}
 After a little bit of algebra, we get
\begin{equation}
F_3=m\frac{2\vert z_1\vert^2+\vert z_2\vert^2}{2 L_1}+
n\frac{\vert z_2-z_1z_3\vert^2-\vert z_3\vert^2}{2 L_2}
\end{equation}
where we defined
\begin{eqnarray}
L_1&=& (1+\vert z_1\vert^2+\vert z_2\vert^2)\nonumber\\
L_2&=&(1+\vert z_3\vert^2+\vert z_2-z_1z_3\vert^2).
\end{eqnarray}
Repeating the same procedure, we get
\begin{eqnarray}
X_8&=&-\frac{i}{2{\sqrt 3}}(z_2\frac{\partial}
{\partial z_2}+z_3\frac{\partial}{\partial z_3})+
\mbox{c.c}
\nonumber\\
F_8&=&\frac{m}{2\sqrt{3}}\frac{\vert z_2\vert^2}{L_1}+
\frac{n}{2\sqrt{3}}\frac{\vert z_2-z_1z_3\vert^2+
\vert z_3\vert^2}{L_2}.
\end{eqnarray}
We can calculate the remaining Hamiltonian functions in a
similar manner and they generate Poisson-Lie algebra
homomorphism \cite{arno}:
\begin{equation}
\{F_a, F_b\}=-f_{ab}^cF_c,
\label{hom}
\end{equation}
from which we deduce
\begin{equation}
\{F_3, F_8 \}=0
\end{equation}
and no other Hamiltonian functions commute with both
$F_3$ and $F_8$.

Then, we have a system of non-commutative
integrability \cite{fome}.
This can be seen from the fact
that
\begin{eqnarray}
F_1&=&-m\frac{ z_1+\bar z_1}{2 L_1}+
n\frac{( z_2-z_1z_3)\bar z_3+
(\bar z_2-\bar z_1\bar z_3) z_3}{2 L_2}
\nonumber\\
F_2&=&mi\frac{ z_1-\bar z_1}{2 L_1}-
ni\frac{( z_2-z_1z_3)\bar z_3-
(\bar z_2-\bar z_1\bar z_3) z_3} {2L_2},
\end{eqnarray}
and $F_3$ and $F_8$ generate  ${\cal G}=su(2)\times u(1)$
 Poisson-Lie algebra
with the property that
\begin{equation}
\mbox{dim}\ {\cal G}+\mbox{rank}\ {\cal G}=6,
\end{equation}
which is equal to $\mbox{dim}\  M_3$.
In this case, the level set
$M_c=\{x\in M:F_i=c_i, \ \ i=1,2,3,8\}$ is a smooth
2-dimensional torus $T^2$ and it is
invariant under $F_3$ and $F_8$ for some $c_i$. Furthermore,
one can find a second commutative algebra ${\cal G}^\prime$
such that
dim ${\cal G}^\prime=3$ \cite{fome}.
It is easy to see that ${\cal G}^\prime$ is generated by
$F_3, F_8$ and the Casimir invariant
$\frac{1}{2}(F_1^2+F_2^2+F_3^2)\equiv C_2$.

The above feature of non-commutative integrability
generalizes to higher flag manifold
$M_N=SU(N)/T^{N-1}$. Consider the generators
of $su(N)$ algebra $T_a$ satisfying
the commutation relations $[T_a, T_b]=-f_{ab}^c T_c$ in
Gell-Mann basis.
Then, repeating the
same procedure as in the case of $su(3)$,
we can calculate all the
Hamiltonian functions $F_a$ associated with each of
the generators $T_a$
and these $F_a$'s satisfy the Poisson-Lie algebra (\ref{hom}).
Obviously, there exist $N-1$ commuting Hamiltonian
functions $F_3,
F_8, F_{15}, \cdots, F_{N^2-1}$ which are far less then $(1/2)
\mbox{dim}\  M_N$ for higher $N$. However, $SU(N-1)\times U(1)$
group actions on $M_N$ satisfy the criteria for
the non-commutative
integrability \cite{fome}:
\begin{equation}
\mbox{dim}\ {\cal G}+\mbox{rank}\ {\cal G}=\mbox{dim}\ M_N.
\end{equation}
The level set $M_{N_c}=\{x\in M_N:F_i=c_i, \ \ T_i\in
su(N-1)\times u(1)\}$
 is a smooth $N-1$ dimensional torus $T^{N-1}$.
Furthermore, we can find a second commutative algebra
${\cal G}^\prime$ such that
dim ${\cal G}^\prime=\frac{1}{2}\mbox{dim}\ M_N$.
The construction
goes as follows: Denote rank $n$ Casimir invariant of
$su(m)$ algebra by $C_n(m)$ \cite{corn}.
 So, for example, $C_2(3)=\frac{1}{2}(F_1^2+
F_2^2+\cdots+F_8^2)$. Then,  ${\cal G}^\prime$ is generated by
$F_3, F_8, \cdots, F_{N^2-1}$,
 $C_p(q)-C_p(q-1),\ p\leq q=2,3,\cdots,N-1$
where we take $C_p(q)=0$ for $p>q$.
For example, in $SU(4)$ case, we have six commuting
functions $F_3, F_8, F_{15}, C_2(2)=1/2(F_1^2+F_2^2+F_3^2),
C_2(3)-C_2(2)=1/2(F_4^2+F_5^2+F_6^2+F_7^2+F_8^2), C_3(3)=
d_{abc}F_aF_bF_c$, where $d_{abc}$ is the symmetric
structure constant of $su(3)$ \cite{corn}.

We consider the integrable system with $F_3, F_8$, and $C_2$
 in involution with
Hamiltonian given by
\begin{equation}
H\equiv H(F_3, F_8, C_2).\label{hami}
\end{equation}
According to the Liouville theorem, we can
find action-angle variables
$(I_1,I_2,I_3, \phi_1,\phi_2,\phi_3)$ such that
the original symplectic
two form Eq.(\ref{symp}) can be expressed as a Darboux form:
\begin{equation}
\Omega=\sum_{i=1}^3 dI_i\wedge d\phi_i
\end{equation}
 They are given by\cite{john}
\begin{equation}
I_1=F_3,\quad I_2=F_8, \quad I_3=\sqrt{C_2}.
\end{equation}
Note that $I_1$ and $I_2$ are global Hamiltonian functions, but
$I_3$ is a local Hamiltonian function. It can not be extended to
the entire $M_3$. In other words, $I_1$ and $I_2$ generate
global torus symplectic actions, whereas $I_3$
generates a local one.

Canonical quantization of the Hamiltonian Eq.(\ref{hami}) could be
rather simple, because the Hamiltonian is diagonalized by
construction.
However, path integral quantization is  non-trivial as was
pointed out in Ref.\cite{niel,alek,john}.
 We perform coherent state path integral
of our integrable system, especially by restricting our
Hamiltonian to be a
linear function of the global torus actions $I_1$ and $I_2$,
\begin{equation}
H_t=\omega_1Q_1+\omega_2Q_2\label{torus}
\end{equation}
where we defined
\begin{eqnarray}
Q_1&=&F_3+\sqrt{3}F_8=
m\frac{\vert z_1\vert^2+\vert z_2\vert^2}{L_1}+
n\frac{\vert z_2-z_1z_3\vert^2}{L_2}\nonumber\\
Q_2&=&F_3-\sqrt{3}F_8=
m\frac{\vert z_1\vert^2}{L_1}-
n\frac{\vert z_3\vert^2}{L_2},
\end{eqnarray}
as a suitable linear combination of $I_1$ and $I_2$ to achieve the
calculational simplicity. $Q_1$ and $Q_2$ generate the following
symplectic torus actions:
\begin{equation}
Q_1: (z_1,z_2,z_3)\mapsto (e^{i\theta_1}z_1,
e^{i\theta_1}z_2, z_3), \quad
Q_2: (z_1,z_2,z_3)\mapsto (e^{i\theta_2}z_1,z_2,  e^{-i\theta_2}z_3).
\end{equation}
This Hamiltonian is special in the sense that the
semiclassical approximation to the path integral
gives the exact
expression for the quantum mechanical propagator due to the
Duistermaat-Heckman (D-H) integration
formula\cite{duit} which found many applications
in physics and mathematics recently \cite{ston,oh,blau}.

To see this, we perform the coherent state path integral
explicitly \cite{klau}. Let us define
\begin{equation}
\vert Z\rangle= \sum_{i=1}^{3}\exp (z_i E_i)\vert 0\rangle
\label{coh}
\end{equation}
where $Z=(z_1,z_2,z_3)$, $E_i$ are the three positive roots and
 $\vert 0\rangle$ is the highest weight vector.
The normalization for Eq.(\ref{coh}) is chosen so that
\begin{equation}
\langle Z^\prime\vert Z\rangle=(1+\bar z_1^\prime z_1+
\bar z_2^\prime z_2)^m(1+\bar z_3^\prime z_3+(\bar z_2^\prime -
\bar z_1^\prime \bar z_3^\prime)( z_2- z_1z_3))^n\label{norm}
\end{equation}
Notice that this definition differs from the usual
one \cite{klau} by
the normalization factor $N=L_1^{-m}L_2^{-n}$.
We have chosen this definition here because in the subsequent
analysis, $\bar Z$ and $Z$
can be treated independently and the over-specification problem
can be side-stepped\cite{klaud,fadd2,oh}.
Then, the resolution of identity is expressed as
\begin{equation}
I=N\int d\mu(\bar Z,Z)  \vert Z\rangle\langle Z\vert\label{reso}
\end{equation}
where $d\mu(\bar Z,Z)$ is the Liouville measure.

Our main interest lies in the evaluation of the propagator
\begin{equation}
K(\bar\xi^{\prime\prime},\xi^\prime;t)
= \langle\bar\xi^{\prime\prime}\vert e^{-i{\hat H}t}\vert
\xi^\prime\rangle.
\end{equation}
Divide the time $T\equiv t^{\prime\prime}-t^\prime=S\epsilon$
into $S$ equal intervals, $Z\equiv Z(p),
\ p=0,1,\cdots,S$ and the boundary condition is given by
\begin{equation}
\bar Z(t^{\prime\prime})=\bar \xi^{\prime\prime}, \quad
Z(t^\prime)=  \xi^\prime.\label{boun}
\end{equation}
Inserting Eq.(\ref{reso}) repeatedly, we have
\begin{equation}
 \langle\bar\xi^{\prime\prime}\vert e^{-i{\hat H}t}\vert
\xi^\prime\rangle=\int \prod_{p=1}^{S-1} d\mu(p)N(p)\prod_{p=1}^S
 \langle Z(p)\vert e^{-i{\hat H}\epsilon}\vert Z(p-1)\rangle
\end{equation}
Using $ e^{-i\epsilon{\hat H}}=I-i\epsilon{\hat H}$ and
the normalization condition
Eq.(\ref{norm}) to evaluate $ \langle Z(p)\vert  Z(p-1)\rangle$,
 we get in the limit
$\epsilon\rightarrow 0$,
\begin{equation}
K=\int d\mu\  \exp\left\{m\log L_1(\bar\xi^{\prime\prime},
Z(t^{\prime\prime}))
+n\log L_2(\bar\xi^{\prime\prime}, Z(t^{\prime\prime}))+
i\int_{t^\prime}^{t^{\prime\prime}} L dt\right\}
\end{equation}
where we defined
\begin{eqnarray}
L_1(\bar\xi^{\prime\prime}, Z(t^{\prime\prime}))&=&
(1+\bar \xi_1^{\prime\prime}
z_1(t^{\prime\prime})+\bar \xi_2^{\prime\prime}z_2(t^{\prime\prime}))
\nonumber\\
L_2(\bar\xi^{\prime\prime}, Z(t^{\prime\prime}))&=&
(1+\bar \xi_3^{\prime\prime} z_3(t^{\prime\prime})+
(\bar \xi_2^{\prime\prime}-\bar \xi_1^{\prime\prime}
\bar \xi_3^{\prime\prime})
( z_2(t^{\prime\prime})-z_1(t^{\prime\prime})
 z_3(t^{\prime\prime}))),
\end{eqnarray}
and the Lagrangian is given by
\begin{equation}
L=mi\frac{z_1\dot{\bar z_1}+ z_2\dot{\bar z_2}}{L_1}+
ni\frac{ z_3\dot{\bar z_3}+( z_2- z_1 z_3)
(\dot{\bar z_2}-\dot{\bar z_1}\bar z_3-\bar z_1\dot{\bar z_3})}{L_2}-H_t
\end{equation}
with $H_t=\omega_1Q_1+\omega_2 Q_2$.

The equations of motions are
\begin{equation}
i\dot z_i=\bar g^{ij}\frac{\partial H_t(\bar Z, Z)}
{\partial \bar z_j},\quad  i\dot {\bar z_i}=-g^{ij}
\frac{\partial  H_t(\bar Z, Z)}{\partial z_j}.
\end{equation}
The solution of equations of motion is linearized completely:
with the boundary conditions
given by Eq.(\ref{boun}), we have
\begin{eqnarray}
z_1(t)&=&\xi_1^\prime e^{i(\omega_1+\omega_2)(t-t^\prime)},\ \ \
z_2 (t)=\xi_2^\prime e^{i\omega_1(t-t^\prime)},\ \ \
z_3(t)=\xi_3^\prime e^{-i\omega_2(t-t^\prime)}\nonumber\\
\bar z_1(t)&=&\bar\xi_1^{\prime\prime}e^{-i(\omega_1+\omega_2)
(t-t^{\prime\prime})},\ \ \
\bar z_2(t)=\bar \xi_2^{\prime\prime}
e^{-i\omega_1(t-t^{\prime\prime})},\ \ \
\bar z_3(t)=\bar \xi_3^{\prime\prime}
e^{i\omega_2(t-t^{\prime\prime})}
\end{eqnarray}
 Denoting the above classical solutions by $\bar Z_{\mbox{c}}$ and
$Z_{\mbox{c}}$ and expanding around the classical solutions
\begin{equation}
\bar Z=\bar Z_{\mbox{c}}+\delta\bar Z,
\quad Z=Z_{\mbox{c}}+\delta Z,
\end{equation}
with boundary conditions $\delta\bar Z(t^{\prime\prime})=
\delta Z(t^{\prime})=0$,
we find the following expression for the propagator:
\begin{equation}
K(\bar\xi^{\prime\prime},\xi^\prime;T)=
a(T)L_1(\bar\xi^{\prime\prime}, Z_{\mbox{c}}(t^{\prime\prime}))^m
L_2(\bar\xi^{\prime\prime}, Z_{\mbox{c}}(t^{\prime\prime}))^n
\exp(iS(\bar Z_{\mbox{c}}, Z_{\mbox{c}},T)).
\label{result}
\end{equation}
Here $a(T)$ is the Van Vleck
determinant\cite{klei}
coming from the Gaussian integration of the
fluctuations $\delta\bar Z$
and $\delta Z$.
Substituting the classical solutions into the above equations
and calculating
the Van Vleck determinant as in Ref.\cite{oh}, we find
\begin{equation}
K=(1+\bar \xi_1^{\prime\prime}\xi_1^{\prime}
e^{i(\omega_1+\omega_2)T}
+\bar \xi_2^{\prime\prime}\xi_2^\prime e^{i\omega_1T})^{m}
(1+\bar \xi_3^{\prime\prime}\xi_3^\prime e^{-i\omega_2T}+
(\bar \xi_2^{\prime\prime}-\bar \xi_1^{\prime\prime}
\bar \xi_3^{\prime\prime})
( \xi_2^\prime-\xi_1^\prime\xi_3^\prime)e^{i\omega_1T})^{n}
\end{equation}
which is guaranteed to be exact due to D-H formula.
The exactness of the above propagator can also be
checked by solving the time-dependent Schr\"odinger equation
set up through the geometric quantization of the Hamiltonian
 (\ref{torus}) \cite{oh95}.
Then, wave function in the anti-holomorphic polarization
at arbitrary time is given by
\begin{equation}
\Psi(\bar \xi^{\prime\prime},t)=\int \frac{d\mu(\bar Z, Z)}
{L_1^{m}L_2^{n}}K(\bar \xi^{\prime\prime},Z;t) \Psi(Z,0)
\end{equation}

In summary, we constructed  integrable spin models on flag manifold
by using the symplectic structure explicitly  given in Bruhat
coordinatization of flag manifold.
We find that the systems are non-commutative integrable
and some of the conserved quantities are given by the
Casimir invariants.
We quantized the systems using the coherent
state path integral technique
and found the exact expression for the propagator
for a special case
where the Hamiltonian is given by the  global
symplectic torus action
on the flag manifold, and semiclassical approximation
gives the exact
results due to the D-H integration formula.
Analysis of higher $N$ case, many spin systems and
geometric quantization of these systems will be discussed
in a forthcoming paper \cite{oh95}.

\acknowledgments
This work is supported by Ministry of Education through
the Research Institute of Basic Science and in part by
the KOSEF through C.T.P. at S.N.U.

\end{document}